\documentclass[journal=jacsat,manuscript=article]{achemso}

\usepackage[version=3]{mhchem} 
\setkeys{acs}{articletitle=true,maxauthors=30}
\author{Elham Rezasoltani}
\affiliation[First University]
{Department of Physics, Imperial College London, SW7 2BW, London, UK}
\email{e.rezasoltani14@imperial.ac.uk}
\author{Jaime Martin}
\affiliation[second University]{Centro de Investigacións Tecnolóxicas, Universidade da Coruña, Campus de Esteiro s/n, 15403 Ferrol, Spain}
\author{Sophia C. Hayes}
\affiliation[Third University]{Department of Chemistry, University of Cyprus,  1678 Nicosia, Cyprus}
\author{Despina Heracleous}
\affiliation[Third University]{Department of Chemistry, University of Cyprus,  1678 Nicosia, Cyprus}
\author{Kyriaki Koumenidou}
\affiliation[Third University]{Department of Chemistry, University of Cyprus,  1678 Nicosia, Cyprus}
\author{Rebeca Hernandez}
\affiliation[Forth University]{Instituto de Ciencia y Tecnología de Polímeros (CSIC) C/ Juan de la Cierva, 3 28006 Madrid, Spain}
\author{Natalie Stingelin}
\affiliation[Fifth University]
{School of Chemical and Biomolecular Engineering
, Georgia Institute of Technology, Atlanta, GA, USA}
\author{Carlos Silva-Acu\~{n}a}
\affiliation[Fifth University]
{School of Physics, Georgia Institute of Technology, Atlanta, GA, USA}
\title[An \textsf{achemso} demo]
  {Formation of a highly ordered red phase in a MEH-PPV: polystyrene pseudogels}

\begin{document}

\begin{abstract}
   In this work, we demonstrate the formation of a  "red-phase" poly[2-methoxy, 5-(2'-ethyl-hexoxy)-1,4-phenylene vinylene-PPV] (MEH-PPV) embedded into a host matrix of highly entangled ultra-high molecular weight polystyrene (MEH-PPV/UHMW PS pseudogel) that allows the simple processing of the MEH-PPV solutions. We processed a "red-phase" in the gel, the gel shows that  the features what have beed demonstrated in the solution can be observed in the processable gel for optoelectronics applications. [Yamagata, Hajime,  and Hestand, Nicholas J.  and Spano, Frank C.  and K\"{o}hler, Anna  and  Scharsich, Christina  and Hoffmann, Sebastian T.  and B\"{a}ssler, Heinz, The Journal of Chemical Physics, 2013, 139, 114903]
\end{abstract}
\section*{Introduction}
The application of quantum optoelectronic devices in both everyday life and advanced science would be widespread; however, the development of high-throughput quantum optoelectronics technologies has proven problematic, due to the absence of easily-scalable methods to process materials of the desired optical and morphological characteristics. Moreover, the solid-state microstructure of polymeric semiconductors can be extremely sensitive to processing conditions. Controlling structure-property relationships of solution-processed conjugated polymers has a strong impact on the intrachain (chain conformation) and interchain (packing) order. In the current manuscript we address that problem by adopting a highly successful approach that will be employed in quantum optoelectronics, which will contribute to the explosion of a new scientific field, and new technologies. Specifically, this work produces a step change for quantum optoelectronics applications through the employment of gel-processed materials and provides an integrated study that aims to harness new fundamental knowledge on the phase-transition exhibited by MEH-PPV through steady-state spectroscopy. The concept of this research was developed by the discovery of "Room-temperature Bose-Einstein condensation of cavity exciton-polaritons in a polymer", by Mahrt et al. in 2014.~\cite{plumhof2014room} Bose-Einstein condensation produces polaritons in microcavities with the prospect for low-threshold organic (plastic) lasers. Hence, we provide new fundamental insights as to the applicability of this polymer in strongly-coupled microcavities. The evidence from the work of K\"{o}hler et al. shows that poly[2-methoxy, 5-(2'-ethyl-hexoxy)-1,4-phenylene vinylene-PPV] (MEH-PPV) solution in a polar solvent environment undergoes a transition from the coil conformation at room temperature (blue phase) to a chain-extended conformation at low temperature (red phase).~\cite{doi:10.1021/ja302408a} K\"{o}hler  et. al. found that the effective conjugation length increases upon cooling the solution consistent with a chain-extended and ordered conformation that characterises a J-aggregated polymer. This property of MEH-PPV makes it a suitable candidate for strongly-coupled microcavities. However, the red-phase of MEH-PPV has only been observed in the solution and there has not been any successful effort to obtain this behaviour in thin film of MEH-PPV. This issue is of particular importance for the development of quantum-optoelectronics devices, which typically incorporate a thin film of the polymer rather than a solution. The significance of this work is to address the current challenge by creating gel-like materials to incorporate MEH-PPV into a polymeric host matrix, which serves as a mechanical support for the chromophores. Thus, MEH-PPV chains experience the same chemical environment as in the conventional solution, while the material is macroscopically gel-like and, thus, it can be processed.

To date, several studies have explored the conformation transition in the thin films of PPV derivatives and MEH-PPV solution.
Zeng et al. measured the PL and Raman spectra of thin films of PPV  using dichloro-p-xylene and tetrahydronthionphene in the temperature ranging from 83\,K to 293\,K. The authors observed no significant shift in the Raman band frequency nor in the PL and they measured the emission energy shift for the $0-0$  electronic transition to be only 6\,meV. The authors also found that the conjugated length is elongated by 2.2 repeat units from 294 to 83\,K and assigned it to the reduced disorder at low temperatures.~\cite{0953-8984-16-28-035} 
Wantz et al. measured electroluminescence of thin film of PPV derivative (BDMO-PPV) in the temperature range of 80-350\,K. The authors found that energy separation (between pure electronic transition $0-0$ and first vibronic transition $0-1$) was temperature independent. The authors also carried out temperature dependent PL measurement of MEH-PPV and found that the $0-0$ pure electronic transition blue-shifted about 60\,meV (as temperature increased). Moreover, the authors found that the Huang-Rhys (HR) factor was  temperature independent and could not assign the blue-shift in the $0-0$ PL peak  to a reduced conjugation length in the thin film of PPV derivative.~\cite{:/content/aip/journal/jap/97/3/10.1063/1.1845580}
 In the work of Silva et al. temperature-dependent PL spectra of thin film MEH-PPV deposited by spin coating on a copper substrate was studied.  The authors showed that $0-0$ pure electronic transition blueshifted about 60\,meV between 13-297\,K. Moreover, Silva et al. found that $0-1$ to $0-0$  peak intensities of the PL spectrum increased as temperature increased and attributed that to the decrease of conjugation length.~\cite{:/content/aip/journal/jcp/128/9/10.1063/1.2835606} 
 Following the work by K\"{o}hler et al. that demonstrated a conformation transition of MEH-PPV in methyltetrahydrofuran (MeTHF),~\cite{doi:10.1021/ja302408a} Yamagata et al. reported a very comprehensive experimental work along with numerical calculation on the MEH-PPV in MeTHF. The authors proposed that the "HJ" hybrid model contributes greatly to the understanding of the basic photophysical features observed for red-phase MEH-PPV.  They concluded that J-like behaviour is favored by a relatively large intrachain exciton bandwidth- roughly an order of magnitude greater than the interchain bandwidth- and the presence of disorder.~\cite{:/content/aip/journal/jcp/139/11/10.1063/1.4819906} 
 
 We introduce here a processed gel-like ultra-high-molecular-weight polystyrene (UHMW PS) polymeric host matrix to trap and protect MEH-PPV solution for use in quantum optoelectronic devices. We demonstrate by means of temperature-dependent absorption and photoluminescence (PL) spectroscopy (ranging from 290\,K down to 80\,K)  that the red-phase feature of MEH-PPV in solution can be retained in MEH-PPV/UHMWPS pseudogels. To this end, we apply the concept of "HJ" aggregate model introduced by Yamagata and et al.~\cite{:/content/aip/journal/jcp/136/18/10.1063/1.4705272}  also applied in MEH-PPV dissolved in methyltetrahydrofuran (MeTHF) by those authors.~\cite{:/content/aip/journal/jcp/139/11/10.1063/1.4819906} We demonstrate that the red-phase of MEH-PPV/UHMW PS pseudogels manifest itself by narrow linewidths and  enhanced $0-0$ line strength in the PL spectrum as well as a small Stokes shifts between the PL and absorption spectra at low temperatures.  Furthermore, the ratio of the 0-0 to 0-1 peak intensities of the PL spectrum of red-phase MEH-PPV/UHMW PS pseudogels is also enhanced and greater than unity relative to the disordered blue-phase, which accounts for the J-aggregation based on the "HJ" hybrid model. We further show that the red phase at low temperatures supports an enhanced coherence length along the chain  compared to the coiled conformation. This highlights that MEH-PPV/UHMWPS pseudogels promote J-aggregate-like behaviour however very disordered. This picture for the  MEH-PPV/UHMW PS pseudogels is consistent with reports by Yamagata et al. on MEH-PPV dissolved in MeTHF.~\cite{:/content/aip/journal/jcp/139/11/10.1063/1.4819906}
\section*{Material processing}
\textbf{Ultra high molecular weight Polystyrene pseudogels (UHMW PS)}
 In this work, the ultra high molecular weight polystyrene ($30\times10^{6}$\,g$/$mol) with molecular weight per repeat unit (M$_{r.u.}$) of  104.1\,g$/$mol (purchased from Polyscience Inc.), has been used as the host matrix system for MEH-PPV solution. (see chemical structure of PS, S1$\dag$)
  It is well known that PS has exhibit thermoreversible gelation at low temperature in a large number of solvents.~\cite{{doi:10.1021/ma00235a006},{Lehsaini19942180}}

  Hence, the first goal was to find the minimum concentration at which the polystyrene solution showed a pseudogel behavior, i.e. absence of macroscale flow within long timescales. In this case, the flow of UHMW PS solutions (in MeTHF) was evaluated for periods of 8-12\,h, thus significantly longer than the timescales of typical processing procedures. To produce the pseudogels first 10, 15, 30, 50, and 100\,mg of polystyrene powder were weighted in separate vials, dissolved in 900\,$\mu$l of Me-THF solvent and stirred for 5-10 minutes  on a hot plate  at a temperature of 80\,$^{\circ}$C, using a heating block for well-defined dissolution temperature. After which the solutions were subsequently heated at a temperature of 58\,$^{\circ}$C (below the boiling point of MeTHF \textit{i.e.} 80\,$^{\circ}$C) for up to one hour - depending on the polystyrene weight fraction - until the solutions were homogenous. The heat was subsequently turned off and the solutions were cooled to room temperature. Upon cooling the solutions to room temperature, polymer chains form a highly dense entanglement network, which behaves like a pseudogel over a range of timescales including those of typical processing methods.  Note that the solution which contained minimum solute (10\, mg of polystyrene), did not appear to form a pseudogel structure. Hence, the lowest concentration of this polystyrene required for the transition from a liquid to a pseudogel is estimated to be 15\,mg$/$ml. As such we obtained the polystyrene matrices at different weight fractions (wt$\%$1.5, 3.0, 5.0, and 10.0) of polystyrene with respect to the total weight of the pseudogels. 

\begin{figure} [ht]
 \centering 
 \includegraphics[width=0.68\textwidth]{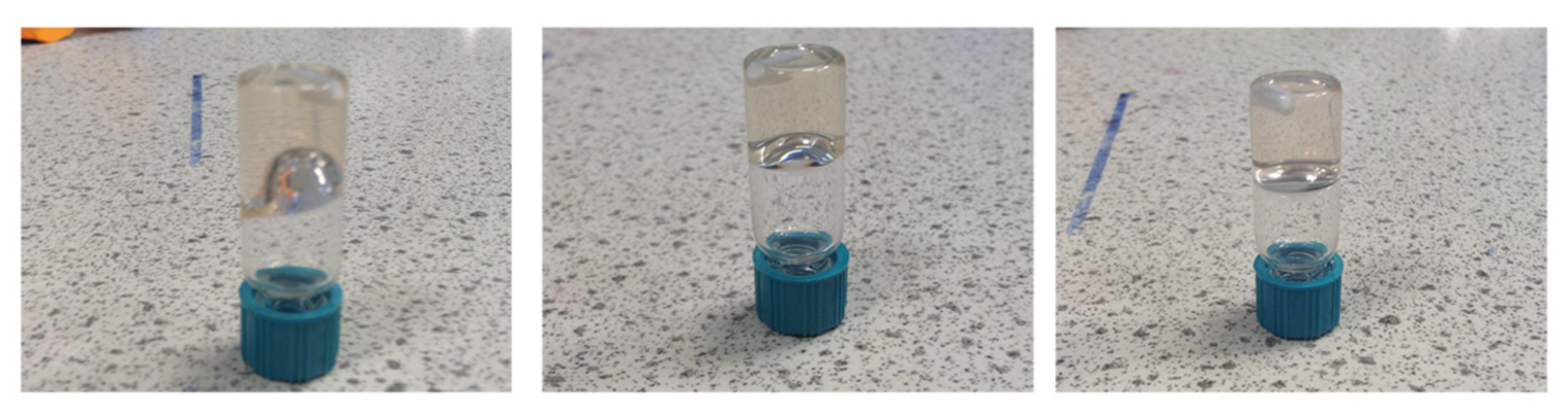}
 \caption{  15\,mg/ml, 30\,mg/ml, 50\,mg/ml U\-H\-M\-W PS pseudogel, solution containing mol (repeating units)$/$lit MEH-PPV. }
 \label{fig:gel}
 \end{figure}

\noindent \textbf{Preparation of  MEH-PPV solution} 
  Similarly, in order to make the MEH-PPV solution  of $5.0\times10^{-3}$\,mol (repeating units)$/$lit,  first 3\,mg of MEH-PPV  powder with a weigh-average molecular weight ($M_{w}$) of $6\times10^{-5}$\,g$/$mol,  with molecular weight per repeat unit (M$_{r.u.}$) of  276\,g$/$mol) was weighed in a vial and then dissolved in 2\,ml of Me-THF solvent which was stirred for 5-10 minutes on a hot plate at 120\,$^{\circ}$C using the heating block for well-defined dissolution temperature. After which the solution was diluted to $5.0\times10^{-4}$\,mol (repeating units)$/$lit, by adding 900\,$\mu$l of Me-THF to the 100\,$\mu$l of the solution. (see chemical structure of MEH-PPV, S1$\dag$)  
 \noindent  \textbf{Ultra high molecular weight polystyrene / MEH-PPV pseudogels}
 MEH-PPV$/$U\-H\-M\-W PS pseudogels at two different molarities of MEH-PPV solution were prepared, by introducing  100\,$\mu$l of  $5.0\times10^{-3}$\,mol (repeating units)$/$lit and $5.0\times10^{-4}$\,mol (repeating units)$/$lit  MEH-PPV solution into the polystyrene pseudogel. The  MEH-PPV$/$U\-H\-M\-W PS pseudogels were heated at 50\,$^{\circ}$C until the MEH-PPV solutions were  thoroughly mixed to obtain the desired homogenous pseudogels. (see tables S2$\dag$) 
 Fig. \ref{fig:gel} illustrates photographs of the MEH-PPV$/$U\-H\-M\-W PS pseudogels for three different concentrations of PS solution (1.5, 3.0, and 5.0 wt.$\%$ ) with $5.0\times10^{-6}$\,mol (repeating units)$/$lit MEH-PPV solution that were subject to a fast cooling rate. 
Because the MEH-PPV solution is too dilute the colour of the pseudogel is pale. The pictures were taken 5 minutes after the vials were held upside down and the pseudogels did not flow downwards. 
 \section*{Experimental methods}
\textbf {Rheology tests}
 Further confirmation on the gel-like behaviour of UHMW PS pseudogels can be obtained from rheological tests. Oscillatory rheological measurements were carried out in a rheometer (AR-G2, TA instruments) using a flat steel plate geometry (40\,mm diameter) equipped with a solvent trap to prevent sample evaporation. Storage modulus  ${G}'$ and loss modulus ${G}''$ were determined in UHMW PS ($M_{w}$=30,000,000\,g/mol) at two different concentrations ( 3.0\,wt$\%$ and 6.0\,wt$\%$) by using an oscillatory frequency sweep test between 0.01 and 100\,Hz at a constant strain of 10\% within the linear viscoelastic region. All measurements were performed at 263\,K.  To be considered a gel, the system must meet some requirements according to its rheological behaviour; (i) the dynamic elastic modulus ${G}'$ must be relatively independent of the frequency of deformation and (ii) ${G}'$ must be greater than ${G}''$ at all frequencies[3]. Fig. \ref{fig:spain} shows the evolution of ${G}'$ and ${G}''$ with frequency for UHMW PS pseudogels at two different PS concentrations ( 3.0\,wt$\%$ and 6.0).\cite{Rebeca}
 
 \noindent\textbf{Temperature-dependent absorption and PL spectroscopy}
Absorption and photoluminescence (PL) spectra were taken from  MEH-PPV$/$U\-H\-M\-W PS pseudogels,  3.0\,wt$\%$ in MeTHF solution at a concentration of  $5.0\times10^{-5}$\,mol (repeating units)$/$lit. We measured the absorption and PL spectra using standard lock-in techniques. For absorption we measured the transmission through the sample  using a home-built setup  consists  of a monochromated 250\,W halogen-tungsten lamp modulated at $f_\textrm {probe}=139$\,Hz and monitored with a Si/Pbs dual-band photo-receiver.  We measured the PL spectrum of the samples using a  continuous-wave laser modulating at $f_\textrm {pump}=170$\,Hz provided the excitation source at wavelength of 405\,nm (3.06\,eV photon energy). The signal was monitored with a Si/avalanche photodiode. The signal was then divided in the $X$ and $Y$ channels of the lock-in amplifier and the phase of the lock-in amplifier was set by placing the scattered pump laser light in the $X$ channel upon modulation at $f_\textrm {pump}$.  The samples in a fused silica 1\,mm cuvette were held in a closed-cycle, sample-in-vapour cryostat,  where it could be cooled down to 10K with exchange gas.

For an analysis of the PL spectra, we need to precisely know the ratio between the intensity of the $0-0$ and $0-1$ peaks in the  PL spectra.  We have modeled  the PL spectrum, $\bar{S}(\hbar \omega)$, as Franck-Condon (FC) transitions involving several intramolecular vibrational modes, $i$, according to~\cite{PhysRevLett.98.206406}
\begin{equation}
\label{equ:FC}
\bar{S}\left ( \hbar \omega \right ) =n^{3}\left ( \hbar \omega \right )^{3}\sum_{m_{i}}\prod_{i}I_{0\rightarrow m_{i}}\times\Gamma \left [\hbar \omega-\left ( \hbar \omega_{0-0}-\sum_{i} m_{i}\hbar \omega_{i}\right )  \right ],
\end{equation}
where $n$ is the refractive index of the material at the optical frequency $\omega$, $\omega_{i}$ is the frequency of the $i$th mode, $m_{i}$= 0,1,2,. . . is the number of vibrational quanta in the $i$th mode, and the set of integers {$m_{i}$} identifies a particular combination peak. $\hbar \omega_{0-0}$ is the $0-0$ peak energy. $\Gamma$ is a Gaussian line shape function with constant standard deviation $\sigma $ that represents the inhomogeneously broadened spectral line of the vibronic replica in the progression. $\hbar \omega_{i}$ is the energy of the effective oscillator coupled to the electronic transition (the most strongly coupled mode (or cluster of modes) corresponds to the high frequency vinyl-stretching mode), $\sim0.174$\,eV for most of the polymeric semiconductors.~\cite{{:/content/aip/journal/jcp/139/11/10.1063/1.4819906},{PhysRevB.88.155202}} The cubic dependence in equation \ref {equ:FC} on the refractive index $n$ of the surrounding medium and on frequency arises from considering the influence of the photon density-of-states in the medium surrounding  its emission coefficient.~\cite{:/content/aip/journal/jcp/139/11/10.1063/1.4819906}
 Furthermore, we  introduce the effective Huang-Rhys (HR) factor from the measured PL spectra. The HR factor is defined as the ratio of the $0-2$ to $0-1$ line strengths given by
 \begin{equation}
 \lambda^{2} =2\frac{I^{0-2}}{I^{0-1}}
 \end{equation}
 where $I^{0-\nu}$ is the line strength corresponding to the $0-\nu$ PL transition.  Paquin et al. in a previous work reported that  the effective HR factor  is far more sensitive to the intrachain exciton
bandwidth than the interchain exciton bandwidth whilst  the $0-0$ to $0-1$ ratio  responds to the exciton coherence number.~\cite{PhysRevB.88.155202}
\textbf{Resonance Raman spectroscopy}
The samples were excited at 435.7\,nm, which was obtained from Raman-shifting in H2 gas the second harmonic of an Nd:YAG laser at 532\,nm (PRO 230-30 Hz, Spectra Physics).  Both solutions and pseudogels were placed in a custom-made cryocell that was fitted to the cold finger of a sample-in-vacuum closed-cycle cryostat (CCS-150, JANIS). The sample cavity had a 13\,mm diameter and 2\,mm pathlength. The cryostat was fitted on a translation stage for periodic translation of the sample during data acquisition. In addition, modest excitation powers (0.23\,mW) were employed to avoid decomposition of the sample. The Raman scattered light was collected in a backscattering geometry and delivered to a 0.75\,m focal-length Czerny-Turner spectrograph, equipped with a 1200-grooves/mm UV-enhanced holographic grating. The slit width was set to 100\,$\mu$m providing for 5\,cm$^{-1}$ spectral resolution. The scattered light was detected by a LN2-cooled 2048$\times$512 pixel, back-illuminated UV-enhanced CCD detector (Spec10:2KBUV/LN, Princeton Instruments).  Each spectrum presented here is the accumulation of 6-10\,min spectra in the case of solutions and between 9-5 10\,min spectra in the case of pseudogels. Raman spectra were obtained at room temperature, 180 and 110\,K. Frequency calibration of the spectra was accomplished with the use of cyclohexane. MATLAB was used for spectral treatment and analysis.
\section{Results and Analysis}
\begin{figure} [h!]
 \centering
 \includegraphics{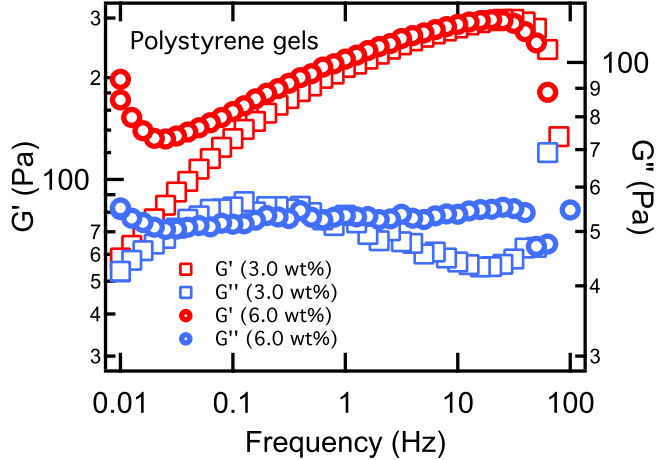}
 \caption{  Frequency sweep on the U\-H\-M\-W PS solution ( 3.0\,wt$\%$ and 6.0\,wt$\%$). The log-log plot of elastic moduli ${G}'$ in red (left axis) and  ${G}''$ in blue (right axis)  as a function of frequency, open squares corresponded to   3.0\,wt$\%$ and filled circles corresponded to   6.0\,wt$\%$ respectively. The more frequency dependent the elastic modulus is, the more fluid-like is the material. }
 \label{fig:spain}
 \end{figure}

Fig. \ref{fig:spain} shows the frequency sweep  at a constant temperature of 10\,$^{\circ}$C and using a deformation of 10$\%$, in UHMW PS ($M_{w}$=30,000,000\,g/mol) at two different concentrations ( 3.0\,wt$\%$ and 6.0\,wt$\%$).
  The storage ${G}'$ modulus (in red - left axis) and the loss modulus  ${G}''$ (in blue - right axis) are plotted as a function of frequency, with open and filled circles corresponding to  6.0\,wt$\%$ and  3.0\,wt$\%$ respectively. According to the frequency sweep done in the two samples, it is clear that ${G}'>{G}''$ in both samples, as expected for a gel-like material.~\cite{Rebeca} Moreover, the results demonstrate  that  the elastic modulus ${G}''$ is nearly independent of frequency, and  ${G}'$  is frequency dependent. This behaviour is more evident for the sample which has lower polystyrene concentration, because it shows a terminal region at low frequencies where ${G}'$ and ${G}''$ decrease with frequency. As such we conclud that the samples are highly viscous liquids.
The absorption and PL spectra were taken from  MEH-PPV$/$U\-H\-M\-W PS pseudogels  3.0\,wt$\%$, MEH-PPV solution containing $5.0\times10^{-5}$\,mol (repeating units)$/$lit. 

\begin{figure}[h!]
  \includegraphics{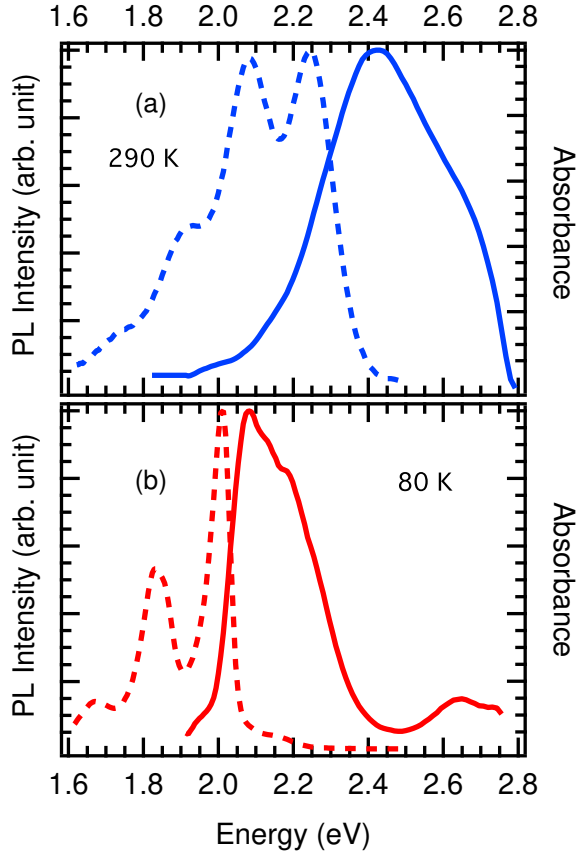}
  \centering
  \caption{Absorption and PL intensity of MEH-PPV/ U\-H\-M\-W PS pseudogels obtained in PS  3.0\,wt$\%$ solution containing $5.0\times10^{-5}$\,mol (repeating units)$/$lit. taken at (a) 290\,K,  where the blue phase dominates, and (b) 80\,K,  where the red phase dominates, as a function of photon energy.}
  \label{fig:PL-gel} 
\end{figure}

We applied temperature-dependent absorption and PL spectroscopy to explore the temperature induced transformation from the blue phase to the red phase. As shown in Fig. \ref{fig:PL-gel}  the absorption spectra at 80\,K and  290\,K  are red-shifted from 2.42\,eV to 2.1\,eV with narrower bandwidth going from  290\, K to 80\,K. 

In addition the $0-0$ (2.1\,eV) peak is dominant over the first side band $0-1$ (2.2\,eV).  
The ratio of the oscillator strengths of the $0-0$ and $0-1$ peaks, R$_{abs}$ $\equiv$ $0-0$ to $0-1$, is 1.25, consistent with R$_{abs}$ of MEH-PPV solution reported by Yamagata et al.~\cite{:/content/aip/journal/jcp/139/11/10.1063/1.4819906} The PL spectrum shows narrower line width with enhanced $0-0$ peak  width at 80\,K compared to 290\,K.   When the pseudogel is cooled the Stokes shift between the PL and the absorption  shrinks  ($\sim67$\,meV) compared to that of at 290\,K. When the absorption and the PL originate roughly from the same energy band, they undergo a small Stokes shift and characterises predominantly J-aggregate behaviour.~\cite{doi:10.1021/ar900233v} 
 In the PL spectra of the pseudogel shown in Fig. \ref{fig:PL-gel} the $0-0$ PL peak intensity corresponds to pure electronic transition and the first ($0-1$) and second ($0-2$) vibronic progression bands are clear at low temperature. 

\begin{figure} [h!]
 \centering 
 \includegraphics{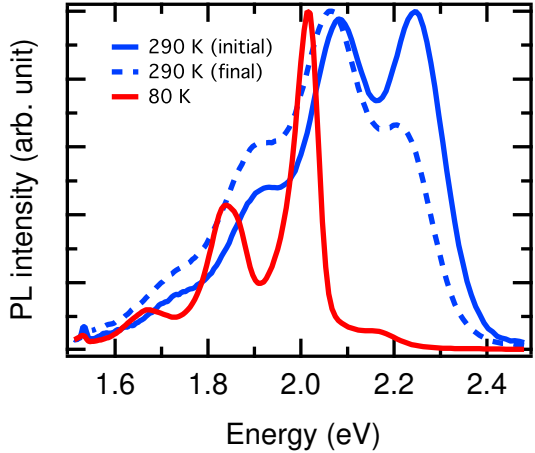}
 \caption{ PL spectra  of  MEH-PPV$/$U\-H\-M\-W PS pseudogels obtained in PS  3.0\,wt$\%$ solution containing $5.0\times10^{-5}$\,mol (repeating units)$/$lit taken at 290\,K (blue solid and dash lines in the cooling and heating cycle respectively) where the blue phase dominates and at  80\,K (in red) where the red phase dominates in the cooling-heating cycle.}
 \label{fig:PL}
 \end{figure}

\begin{figure}[h!]
\centering
 \includegraphics{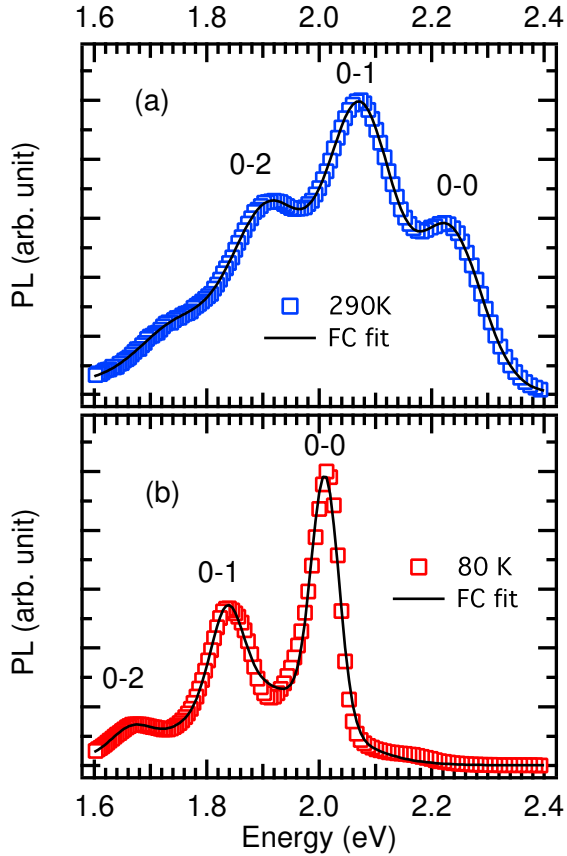}
 \caption{ Franck-Condon analysis to the PL spectra at 290\,K (a) and at 80\,K (b) for  MEH-PPV$/$U\-H\-M\-W PS pseudogels  3.0\,wt$\%$ solution containing $5.0\times10^{-5}$\,mol (repeating units)$/$lit. The PL spectra (circles), and  the solid line shows the Franck-Condon fit. The $0-0$, $0-1$ and the $0-2$ peaks are shown. The PL spectra were divided by the cubic dependence of the radiative decay rate on photon frequency ($\omega^{3}$).}
 \label{fig:FC}
 \end{figure}

\begin{figure}[h!]
  \centering
 \includegraphics{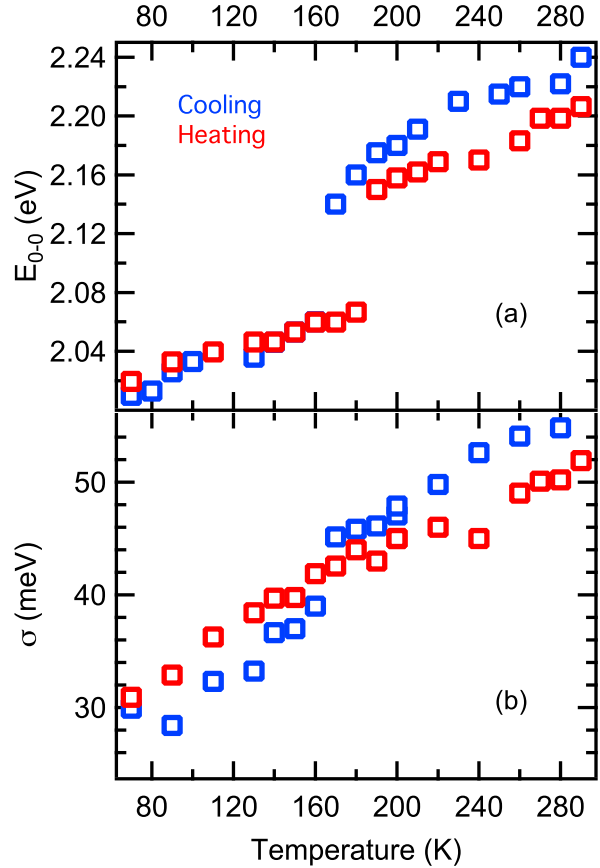}
 \caption{ (a) The positions of the $0-0$ peaks for PL spectra  from the Franck-Condon fits, cooling half-cycle in red and heating half-cycle in blue. (b) the standard deviation $\sigma$ of the Gaussian linewidth for the $0-0$ S$_{0}\rightarrow$S$_{1}$, in U\-H\-M\-W PS/MEH-PPV pseudogels (wt$\%$3.0 solution containing $5.0\times10^{-5}$\,mol (repeating units)$/$lit).}
 \label{fig:Gauss}
 \end{figure}

\begin{figure}[h!]
 \centering
 \includegraphics{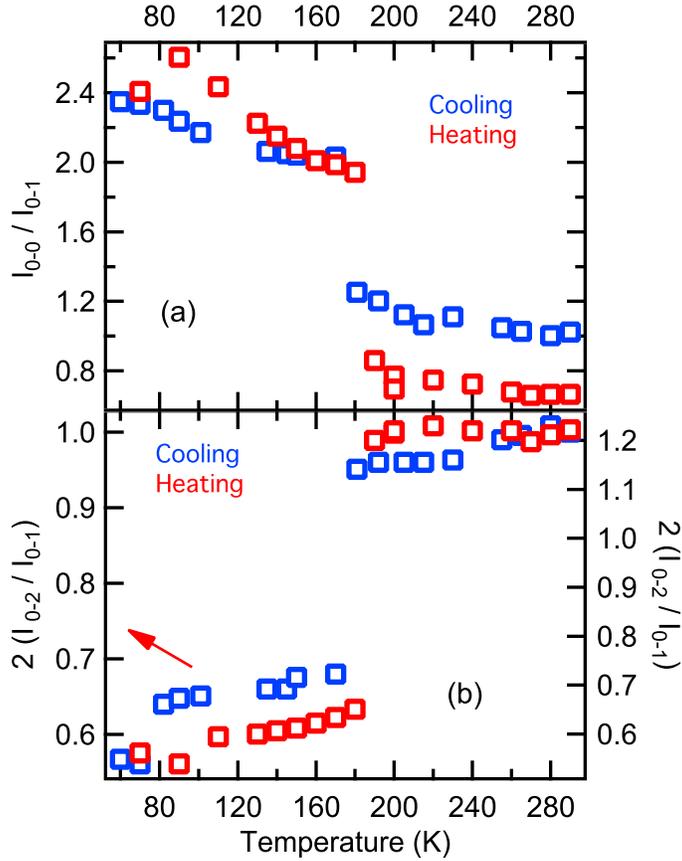}
 \caption{ The ($0-0$ to $0-1$) PL line strength  ratio  as a function of temperature extracted from Franck-Condon analysis  (a) cooling half-cycle in red and heating half-cycle in blue. The $0-0$ to $0-1$ ratio decreases as the temperature grows, that is the characteristic of a predominantly J-aggregate polymer. (b) (2$\times 0-2/0-1$) as a function of temperature extracted from Franck-Condon fitting, cooling half-cycle  in red  and heating half-cycle in blue. The Huang-Rhys parameter increases as temperature increases, that is the characteristic of a predominantly J-aggregate polymer.}
 \label{fig:ratio}
 \end{figure}

\begin{figure}[h!]
 \centering
 \includegraphics{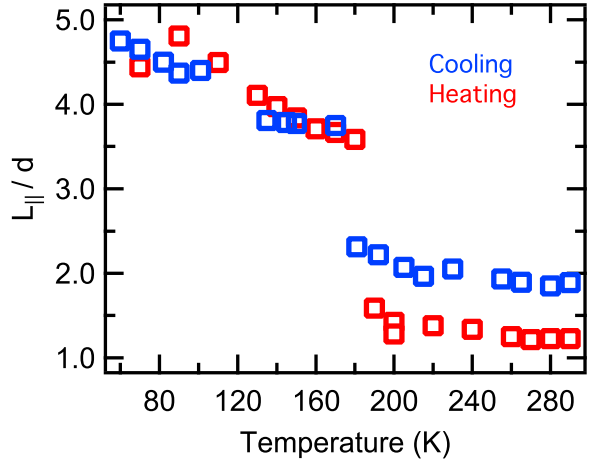}
 \caption{ Exciton coherent length in the cooling-heating cycle for disordered  MEH-PPV$/$U\-H\-M\-W PS pseudogels (wt$\%$3.0 solution containing $5.0\times10^{-5}$\,mol (repeating units)$/$lit). $N_{coh}\sim6$.}
 \label{fig:coherence}
 \end{figure}

\begin{figure}[h!]
 \centering
 \includegraphics{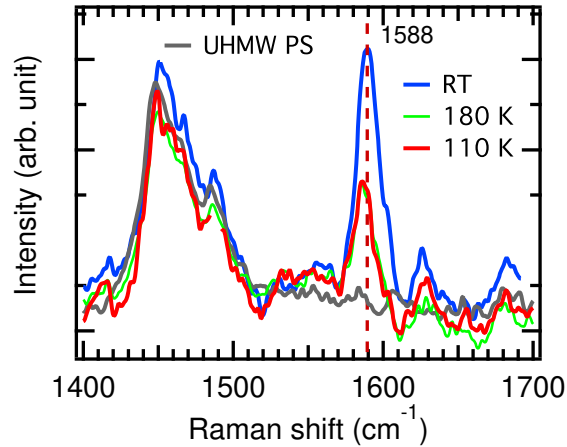}
 \caption{ Temperature dependent resonance Raman spectra of MEH-PPV in UHMW PS/MeTHF pseudogels. The spectra are normalised with respect to the 1448\,cm$^{-1}$ band of MeTHF. The dashed lines indicate spectral changes.}
 \label{fig:gel-raman}
 \end{figure}

 We explore below in more detail  the dependence of  the PL spectra with temperature within a cooling-heating cycle. Aggregation in the molecules cause  the absorption and PL line shapes to change; these changes are reflected  in energy (spectral) shifts and a redistribution of vibronic peak intensities.~\cite{spano2014h} Fig. \ref{fig:PL}, shows that  at  290\,K (initial) in the cooling half-cycle  the intensity of $0-0$ band is nearly equal to that of $0-1$ peak. This is characteristic of  an isolated molecule with where the $0-0$ and $0-1$ line strengths are usually identical. In marked contrast, at low temperature  the $0-0$ peak is enhanced compared to the $0-1$, with the $0-0$ to $0-1$ PL line-strength ratio greater than unity; this is the signature of J-aggregate-like behaviour.~\cite{{doi:10.1021/ar900233v},{spano2014h}} On the other hand, the heating half-cycle shows that by increasing the temperature to 290\,K (final) the spectrum demonstrates an H-aggregate-like behaviour due to the suppressed $0-0$ to $0-1$ PL line-strength ratio to less than unity. 
It is important to note that by repeating the cooling half-cycle in the same sample  the red-phase and the J-like spectrum was retained.
We assess the difference in vibrational peak intensities using a Franck-Condon analysis. 
 The $0-n$ vibronic peak intensities are affected differently by aggregation,~\cite{{:/content/aip/journal/jcp/122/23/10.1063/1.1914768},{PhysRevLett.98.206406}} as for example the attenuated $0-0$ peak in the PL spectrum of H-aggregates. Fig. \ref{fig:FC}(a) and \ref{fig:FC}(b) show the normalised PL spectra of the blue phase (290\,K) and red phase (80\,K), respectively, along with the Franck-Condon fit using equation \ref{equ:FC}. We corrected the PL spectra  for the cubic dependence of the radiative decay rate on photon frequency. We observe a $0-0$(2.01\,eV)$/0-1$(1.83\,eV) PL line-strength ratio that is significantly larger in the MEH-PPV/UHMW PS pseudogels at low temperature compare to a $0-0$ (2.22\,eV))$/0-1$ (2.07\,eV) ratio at 290\,K. 
 Fig. \ref{fig:Gauss}(a) and \ref{fig:Gauss}(b) show the temperature dependence of the pure electronic transition peak energy and the Gaussian line width \textit{i.e.} the standard deviation $\sigma$ of the Gaussian function, extracted  by fitting the PL data to equation \ref{equ:FC}. The redshift of the PL peak energy with decreasing temperature could be associated and to the more planar conformation accompanying chain elongation and increasing effective conjugation length.~\cite{{:/content/aip/journal/jcp/128/9/10.1063/1.2835606},{PhysRevB.44.8652}} 
  Furthermore the peak energy positions, in cooling and heating, reflect a consistent PL shift up to approximately 160\,K above which we can clearly see features corresponding to a hysteresis behaviour. 
 At lower temperatures, $\sigma$ is 30\,meV, which is approximately half of that at high temperatures (\textit{i.e.} 54\,meV in the cooling half-cycle and 51\,meV in the heating half-cycle). The change in $\sigma$ is possibly a consequence of  inhomogeneous line widths and disordered distributions at different temperatures, see Fig. \ref{fig:Gauss}(b).
 The evolution of the  $0-0$ to $0-1$ PL line-strength ratio  as a function of temperature via a fit with the Franck-Condon model (equation \ref{equ:FC}) for the cooling-heating cycle is shown in Fig. \ref{fig:ratio}(a). 
The $0-0$ to $0-1$ PL line-strength ratio is roughly 2-3 at low temperatures  and decreases with increasing temperature. Furthermore, Fig. \ref{fig:ratio}(a)  shows that the temperature dependence of $0-0$ to $0-1$ PL line-strength ratio is not linear and turns over at approximately 170\,K and 180\,K in the cooling and heating half-cycle respectively. 
 The ratio greater than one, also decreasing with increasing temperature characterises a dominant intra-chain (J-like) behaviour.~\cite{spano2014h}  
It is also possible to extract HR factors via a fit with the Franck-Condon model by simply taking the area of $0-1$ and $0-2$ from the measured PL intensity ranging from 80\,K to 290\,K, see Fig. \ref{fig:ratio}(b).  
We deduced values between $\sim0.6-1$ in the cooling half-cycle and values between $\sim0.6-1.2$ in the heating half-cycle. HR factors  are constant over the entire temperature range investigated for the red phase \textit{i.e.} 80\,K to 160-180\,K. Franck-Condon analysis results in values in a good agreement with  values reported by Yamagata et al.  for MEH-PPV ($5.0\times10^{-5}$\,mol (repeating units)$/$lit) solution  within the temperature ranging from 80\,K to 150\,K.~\cite{:/content/aip/journal/jcp/139/11/10.1063/1.4819906} Moreover, HR factors  are constant over the entire temperature range investigated for the blue phase $\sim160-290$\,K. As shown in Fig. \ref{fig:ratio}(b), the values found for HR factors  changes suddenly at the temperature where the red-blue transition occurs from smaller values at low temperatures $\sim0.6$ compared to $\sim1-1.2$ at higher temperature. That can be possibly a consequence of different conformation behaviour in the polymer chains. 
  We can use the measured PL intensities to further evaluate the exciton coherence  length in  MEH-PPV$/$U\-H\-M\-W PS pseudogels. When an exciton is delocalised coherently over a length $L_{C}$ within its aggregate, the exciton wave function is conserved over $L_{C}$.  As such development of  a large coherence length is highly desired since lack of coherence is mainly caused by disorder. To this end, we refer to previous work by Yamagata et al.~\cite{:/content/aip/journal/jcp/139/11/10.1063/1.4819906} on MEH-PPV solution, where the authors showed that  in a J-aggregates exciton localisation behaves similar to the PL intensity ratio, such that the ratio decreases as the temperature rises (and/or disorder increases) which is different from the H-aggregates in which the $0-0$ to $0-1$ increases with increasing temperature (and/or disorder increases).
 Therefor the coherence length in a J-aggregate,  is  given by  
\begin{equation}
 \frac{I_{0-0}}{I_{0-1}}=k \frac{N_{coh}}{\lambda ^{2}},
 \label{equ:coherent}
 \end{equation}
 where $k$ is a prefactor close to unity.~\cite{{:/content/aip/journal/jcp/139/11/10.1063/1.4819906},{:/content/aip/journal/jcp/130/7/10.1063/1.3076079}}
The coherence number at $T=0$\,K in aggregates with no disorder  corresponds to $L_{c}=\left ( N_{coh} -1\right )d$, where $d$ is the distance between nearest-neighbour chromophores within the chain.
By adopting $k=0.62$, $0-0$ to $0-1$ (same as in Fig. \ref{fig:ratio}(a)) and approximate value of unity for $\lambda^{2}$ from our results shown in figure \ref{fig:ratio}(b) and according to $L_{C}=(N_{coh}-1)d$, we found $N_{coh}\sim6$ (see calculation in S3$\dag$). Fig. \ref{fig:coherence} shows the number of coherently connected chromophore as a function of temperature in the cooling-heating cycle. It is clearly shown that  the exciton coherence decreases as the temperature increases, however the exciton becomes even more localised in the heating half-cycle. 
Structural changes in the polymer induced by phase changes were probed as a function of temperature using resonance Raman spectroscopy. The polymer was investigated both in solution and in the UHMW PS pseudogels, in order to assess the extent of structural change in the two environments.  
\begin{figure}[h!]
\centering
 \includegraphics{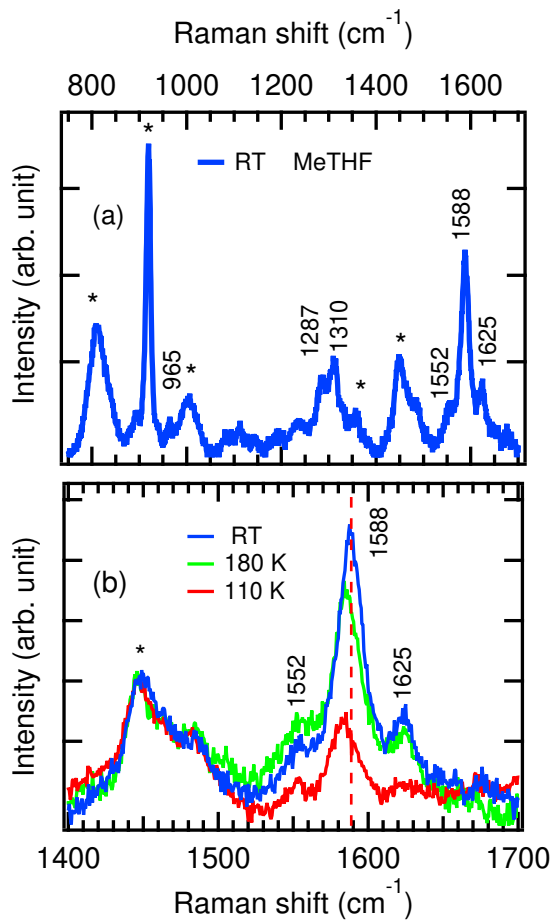}
 \caption{(a) Room temperature resonance Raman spectra of MEH-PPV in MeTHF (5$\times10^{-5}$\,mol (repeating units)$/$lit) excited at 435.7\,nm. The asterisks denote solvent bands. (b) Temperature dependent resonance Raman spectra normalised with respect to the 1448\,cm$^{-1}$ band of MeTHF. The dashed lines indicate spectral changes.}
 \label{fig:sol-raman}
 \end{figure}
The room temperature resonance Raman spectrum of dilute MEH-PPV solutions in MeTHF with excitation at 435.7\,nm is presented in Fig. \ref{fig:sol-raman} (a). The most prominent polymer bands appear at 1588, 1623, 1310, 1286, and 965\,cm$^{-1}$. These correspond to phenyl ring symmetric C=C stretching, vinyl C=C stretching, vinyl C=C-H bending, phenyl ring C=C-H bending, and vinyl C-H out-of-plane bending, respectively.~\cite{{doi:10.1021/nn700213t},{:/content/aip/journal/jcp/127/10/10.1063/1.2767266}} The weak band at 1552\,cm$^{-1}$ corresponds to the asymmetric C=C stretching vibration of the phenyl ring. The temperature dependent resonance Raman spectra of MEH-PPV in MeTHF appear in Fig. \ref{fig:sol-raman} (b), focusing on the region with the most noticeable spectral changes. The temperatures chosen represent the three regimes where the polymer goes from the blue phase (RT), through a transition state (180\,K) to the red phase (110\,K). 
 We observe that decreasing the temperature of the sample leads to a decrease in the intensity of the polymer bands with respect to the multiple solvent bands (bands at 810, 920, 1003, 1354, and 1448\,cm$^{-1}$)~\cite{JRS:JRS1250010103} due to red-shifting of the absorption spectrum leading to significantly decreased resonance enhancement. In addition, the phenyl ring C=C symmetric stretch band downshifts from 1588 to 1583\,cm$^{-1}$ at 110\,K, while the relative intensity between this band and the vinylic C=C stretch increases. Both of these observations are consistent with increased planarisation of the polymer backbone and delocalization of the $\pi$ electron cloud.~\cite{{C2CP41748K},{:/content/aip/journal/jcp/127/10/10.1063/1.2767266},{doi:10.1021/j100182a085}} Another indication for a tendency towards planarity would be a decreasing 965\,cm$^{-1}$  band intensity.~\cite{{:/content/aip/journal/jcp/127/10/10.1063/1.2767266},{doi:10.1021/j100182a085}}  Unfortunately, the small RR cross section for this mode, the reduction in resonance enhancement at 110\,K, along with the small concentration of the polymer in solution and the increased solvent band intensities hindered an unequivocal identification of this mode.  Overlapping solvent bands with the 1287 and 1310\,cm$^{-1}$ modes interfere with the relative intensities of these modes, which have been used in the past to identify conformational changes. 
 The temperature dependent resonance Raman spectra of the polymer in the UHMW PS pseudogels matrix are presented in Fig. \ref{fig:gel-raman} (see the whole range RR spectrum in S4$\dag$). 
 \noindent 
 Very similar RR spectra to the case in solution are obtained in the gel matrix, with the bands associated essentially with the polymer and MeTHF. No polystyrene bands are enhanced at this excitation wavelength. The temperature dependent resonance Raman spectra of the polymer in the UHMW PS pseudogels matrix are presented in Fig. \ref{fig:gel-raman}. Decreased intensity of the bands with temperature is also observed in the case of gels. However, only a 2\,cm$^{-1}$  downshift is observed here in the 1588\,cm$^{-1}$ mode, indicating that either the conformational changes in the gel are limited, or that at 110\,K a mixture of phases is present, with the RR spectrum at 110\,K indicating an average situation. 
 \section*{Discussion}
MEH-PPV in a polar solvent environment MeTHF solution  in the polymeric host matrix shows a conformation transition from disordered coil-like conformation, referred to as a blue phase, and extended-chain-segment conformation, referred to as a red phase. 

The objective of this work was to achieve solution-processable materials and to ensure that the red-phase MEH-PPV is attained in the pseudogel similar to the solution.~\cite{:/content/aip/journal/jcp/139/11/10.1063/1.4819906}
  With the aids of photophysics features (by temperature-dependent PL measurement), we demonstrated identical conformation transition in the  MEH-PPV$/$U\-H\-M\-W PS pseudogels to the corresponding MEH-PPV solution, consistent with the previous works.~\cite{{:/content/aip/journal/jcp/139/11/10.1063/1.4819906},{doi:10.1021/ja302408a}} In addition to the previous works on the MEH-PPV solution, we demonstrated how the conformation changed through a heating half-cycle.
  
 The effects of solvent and environment on PL spectra can be due to several factors in addition to solvent polarity. However since in a particular measurement more than one factor effect can affect the chromophore, it is challenging to know which effect is dominant. 
Given that chromophores dipole moment $\mu_{E}$ is larger in the excited state than in the ground state $\mu_{G}$, following excitation, the polar solvent dipoles will reorient around $\mu_{E}$, which lowers the energy of the excited state. As the solvent polarity is increased, this effect
becomes larger, resulting in emission at lower energies or longer wavelengths.
Moreover, a poor solvent may cause aggregation in the polymer, as the polymer tends to minimise the interaction with poor solvent.~\cite{{opac-b1130600},{doi:10.1021/ma001354d}}

Collison et al.  showed the PL decays for MEH-PPV solutions, of equal concentration but varying solvent quality. The decay of good solvent solutions are almost mono-exponential, with the decay time increasing with decreasing the solvent quality. Collison et al. hypothesised that  this long-lived PL is associated with the singlet state following back-transfer from non-emissive interchain excited states to intrachain excited states. Given that, it is likely that a poor solvent causes the torsional motion along the backbone to be suppressed by the presence of adjacent chains, so that to avoid polymer-solvent interactions. Solvent-induced packing thus contributes to ordering of the chains.~\cite{doi:10.1021/ma001354d}
Furthermore,  Collison et al. explained the room-temperature PL spectral line shape of MEH-PPV dissolved in a toluene-hexane mixture via a two-emitter model and concluded that the spectral line shapes are consistent with single chain; however, the red-phase corresponds to aggregates.~\cite{{doi:10.1021/ma001354d},{:/content/aip/journal/jcp/139/11/10.1063/1.4819906}} In short, the poor solvent can also be a cause for aggregation of MEH-PPV in the pseudogel.

 Moreover, we assign the different conformation transition of the pseudogels -through the cooling-heating cycle- to the physical aggregation of molecules. It is important to note that we prepared a well dissolved diluted solution  at high temperature (120\,$^{\circ}$C), hence at the beginning of the cycle the polymer chains are well separated  from each others. However, as decreasing the temperature  the solubility of the polymer decreases and the polymer chains tend to aggregate especially in poor solvent environment.  This aggregation may occur when the red phase has been formed, therefore, it might not affect the amount of the red phase, meaning that the already planarised molecules might aggregate. 
Then, in the heating half-cycle  this molecular aggregates can be insoluble, and MEH-PPV chains do not show  the behavior of an isolated molecule any longer. 

 Here we discuss the red-phase of the  MEH-PPV$/$U\-H\-M\-W PS pseudogels. We assess PL vibrational peak intensities using Franck-Condon analysis and interpreting the results in terms of the disordered HJ-aggregate model developed  by Ymagata et al.~\cite{:/content/aip/journal/jcp/136/18/10.1063/1.4705272}  

Upon cooling the pseudogel the PL spectra is dramatically ($\sim230$\,meV) red-shifted compared to previous works.~\cite{{:/content/aip/journal/jap/97/3/10.1063/1.1845580},{0953-8984-16-28-035},{:/content/aip/journal/jcp/128/9/10.1063/1.2835606}} PL is redshifted from 2.24\,eV ($0-0$) at 290\,K (initial) to 2.08\,eV ($0-0$) at 80\,K and a reversible blue-shift is observed towards 2.20\,eV at 290\,K (final) when the pseudogel is warmed up (see Fig. \ref{fig:PL}). Yamagata et al. also reported a dramatic redshift ($\sim180$\,meV) in the pure $0-0$ transition of PL spectra from 300 to 80\,K. However K\"{o}hler et al. reported a redshifted PL peak approximately by 130\,meV from 300 to 80\,K.~\cite{doi:10.1021/ja302408a}

In our work, the ratio of the $0-0$ to $0-1$ peak intensities in the  PL spectrum diminishes as temperature rises (see Fig. \ref{fig:ratio}) that is assigned to a J-behaviour.~\cite{:/content/aip/journal/jcp/136/18/10.1063/1.4705272}
Moreover, given that the $0-0$ is enhanced compared to the $0-1$ in a predominantly J-aggregate, a ratio of greater than 1 is expected which is opposite in a predominantly H-aggregate case.  $0-0$ to $0-1$ increases from 1.02 at 290\,K (initial) to 2.38 at 80\,K in the red-phase during the cooling, however dropping from 2.38 at 80\,K to 0.66 at 290\,K (final) during heating half-cycle. It is concluded that the  MEH-PPV$/$U\-H\-M\-W PS pseudogels remains H-aggregated at the end of the cycle. There is also a  broadening of the linewidth, $\sim30$\,meV to $\sim55$\,meV in good agreement with measured values in MEH-PPV solution as reported by Yamagata et al., 24\,meV at 80\,K and 55\,meV at 300\,K.~\cite{:/content/aip/journal/jcp/139/11/10.1063/1.4819906}  
In short, there is evidences of  a predominantly J-aggregation in the red-phase and a predominantly H-aggregation in the blue-phase, which is consistent with the chain-extended conformation to coil conformation mentioned by K\"{o}hler et al.~\cite{doi:10.1021/ja302408a} 
Yamagata et al. hypothesised that the photophysics of MEH-PPV solution results mainly from intrachain $\pi$-electron coupling. Enhanced $0-0$ to $0-1$ at low temperature is also due to the reduced interchain coupling  that can be assigned to the reduced torsional disorder and a higher fraction of intramolecular interactions that contributes to the emission. In addition, thermal disorder is reduced as temperature decreases hence the spatial extent of the electronic wave function increases such that excitons are more delocalised.~\cite{:/content/aip/journal/jcp/139/11/10.1063/1.4819906} 
We further conclude that the smaller values of the HR factor at low temperature is associated with a longer coherence length within the chains, resulting from a more planar configuration and a more extended chain at low temperatures as suggested by the downshift of the phenyl ring C=C stretch in the RR spectra. Furthermore, red-phase MEH-PPV/UHMW PS pseudogel is thus assigned to a disordered J-aggregate compared to the linear J-aggregate PDA in which the $0-0$ PL transition manifests itself by an  ultra-narrow line width as reported in previous works.~\cite{{spano2014h},{:/content/aip/journal/jcp/139/11/10.1063/1.4819906}}  We can conclude that  MEH-PPV$/$U\-H\-M\-W PS pseudogel is a very disorder J-aggregate. 
The results of  MEH-PPV$/$U\-H\-M\-W PS pseudogels are consistent with previous works on the solution of MEH-PPV reported by K\"{o}hler et al. and Yamagata et al.~\cite{{doi:10.1021/ja302408a},{:/content/aip/journal/jcp/139/11/10.1063/1.4819906}} and the red-phase of the solution was successfully reproduced in a host matrix. 

In what follows, we briefly discuss the blue versus the red phase in MEH-PPV/ UHMW PS pseudogels in terms of the recent elegant work developed by Yamagata et al.~\cite{doi:10.1021/jp509011u} and in a later work by Hestand et al.~\cite{:/content/aip/journal/jcp/143/24/10.1063/1.4938012} In the latter work, the authors  discuss hybrid coupling by using a Holstein-style Hamiltonian including both Frenkel and charge-transfer (CT) excitons, where they consider the interference between the two couplings when defining H- and J-aggregates based on short- and long-range couplings. The sign of the CT-mediated coupling is  highly sensitive to small (sub-Angstrom) transverse displacements or slips between chromophores. Moreover, they showed that the interference between short-range and long-range couplings defines four aggregate types: HH, HJ, JH and JJ, based on the sign of the couplings. However, in order to precisely interpret our results and molecule aggregation in  MEH-PPV$/$U\-H\-M\-W PS pseudogels based on this theory, the overlap integrals and coupling strength should be precisely calculated. 
In Fig. \ref{fig:PL}  the solid-blue line in  the PL spectra, 290\,K (initial), demonstrates the beginning of the heating half-cycle. In terms of the developed theory as the strength of the charge transfer and Coulombic couplings are the same, those couplings interfere destructively and create a "null aggregate" which spectroscopically resembles uncoupled molecules.~\cite{:/content/aip/journal/jcp/143/24/10.1063/1.4938012} As such, it is possible to attribute the same $0-0$ and $0-1$ ratio in the  PL spectrum (290\,K (initial)) to the "null aggregate".
We can further assign the red line spectrum in (PL at 80\,K) Fig. \ref{fig:PL}, to either a CT-J or JH aggregate depending on the signs of $J_{Coul}$ and $J_{CT}$. In case of $J_{Coul}=0$ the red-phase PL spectrum reflects a CT-J aggregate. Finally, based on similar argument the PL spectra, 290\,K (final) in Fig. \ref{fig:PL} can be assigned to a CT-H or HJ aggregate.~\cite{:/content/aip/journal/jcp/143/24/10.1063/1.4938012}
It is important to note that in the Kasha limit that  $\left | J_{Coul} \right |\gg \left | J_{CT} \right |$ one can neglect the second letter which corresponds to $J_{CT}$ in for instance JH, giving the conventional "J-aggregate" notation. Therefore, based on this recent theory we come to a similar conclusion for aggregation in  MEH-PPV$/$U\-H\-M\-W PS pseudogels.

In short, this work demonstrates that the morphology of MEH-PPV polymer can be manipulated through altering the critical temperature through which the conformation transition occurs; above the critical temperature the polymer exists in a disordered coil morphology, whilst upon cooing the pseudogel, through the transition temperature, the polymer assembles into aggregates.  The chain-extended conformation or elongated chain in the red phase causes the absorption and the PL spectra to be red-shifted. 

\section*{Conclusion}
We have processed red-phase MEH-PPV in ultra high molecular weight Polystyrene pseudogels.  We demonstrated that the gel-processed U\-H\-M\-W \-P\-S using  a polar organic solvent (MeTHF) with MEH-PPV can lead to a gel-like red-phase of MEH-PPV. We have further implemented temperature-dependent absorption and PL to explore red-blue phase conformation transition in MEH-PPV/UHMW PS pseudogels. We found a dramatic red-shift in, and small Stokes shift  between the absorption and PL spectra with decreasing the temperature  (ranging from 290\,K to 80\,K). Moreover, PL spectra manifest itself by enhanced $0-0$ to $0-1$, and narrow line-width that characterise J-like behaviour based on "HJ" hybrid model.  We note that the substantial increase of $0-0$ to $0-1$ PL peak ratio occurs when MEH-PPV/UHMW PS pseudogels undergo blue-red phase conformation transition.  By Franck-Condon analysis of the PL spectra, we obtained the exciton coherence length ($N_{coh}\sim6$ repeat units). We observed enhanced exciton coherence length at low temperature (where the red-phase dominates) compared to that  at high temperature (where the blue-phase dominates). We  conclude that the J-like behaviour in red-phase MEH-PPV/UHMW PS pseudogels is assigned to reduced torsional disorder and significant chain elongation as confirmed from structural studies demonstrating a downshift of the phenyl ring C=C stretch in the RR spectra. Finally, our observations demonstrate that MEH-PPV/UHMW PS pseudogels undergo coil to chain-extended conformation transition with decreasing the temperature therefore, the red-phase MEH-PPV is retained in the pseudogels.

\bibliography{MEHPPV}
\end{document}


\newpage
\textbf{Molecular structure of poly [2-methoxy, 5-(2'-ethyl-hexoxy)-1,4-phenylene vinylene-PPV] (MEH-PPV) and ultra high molecular weight Polystyrene (UHMW PS) }
\begin{figure}[h!]
\centering
\begin{minipage}{.5\textwidth}
  \centering
  \renewcommand{\thefigure}{S\arabic{figure}}
  \includegraphics[width=.4\linewidth]{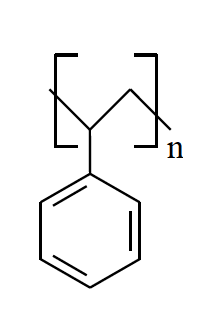}
  \caption{ Molecular structure of PS}
  \label{fig:test1}
\end{minipage}%
\begin{minipage}{.5\textwidth}
  \centering
  \renewcommand{\thefigure}{S\arabic{figure}}
  \includegraphics[width=.4\linewidth]{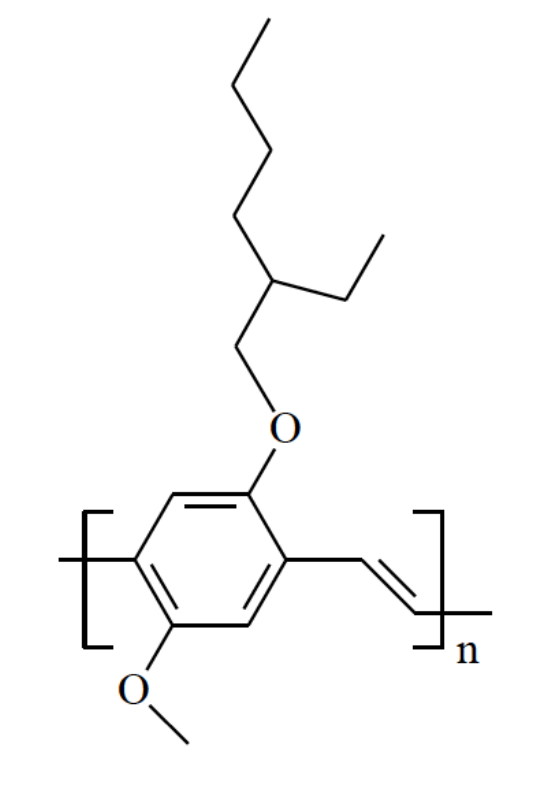}
  \caption{ Molecular structure of MEH-PPV}
  \label{fig:test2}
\end{minipage}
\end{figure}

\textbf{ Weight and mole percent of the Polystyrene solution}
\begin{table}[!ht]
\renewcommand{\thetable}{S\arabic{table}}
\centering
\begin{tabular}{ccc}
  &Polystyrene / MeTHF&\\
 \hline
   Concentration(mg/ml)  & wt.$\%$ &(mole percent$\%$)$\times 10^{-3}$ \\
  \hline
  15& 1.5& 0.15\\
  30&3 & 0.3 \\
  50&5&0.50\\
  100&10 &1.03\\
  \hline
\end{tabular}
\caption{Weight and mole percent of the Polystyrene solution}
\end{table}
\newpage
\textbf{Molarity of  MEH-PPV solution}
\begin{table}[!ht]
\renewcommand{\thetable}{S\arabic{table}}
\centering
\begin{tabular}{ccc}
  &MEH-PPV / MeTHF&\\
 \hline
   Molarity (M)&wt.$\%$&mole percent$\%$ \\
  \hline
  {5.0$\times10^{-5}$\,M}&1.61$\times10^{-4}$&0.62$\times10^{-5}$\\
  {5.0$\times10^{-4}$\,M}&1.61$\times10^{-3}$&0.62$\times10^{-4}$ \\
  \hline
\end{tabular}
\caption{Molarity of  MEH-PPV solution}
\end{table}
\textbf{Characteristics of different  gels}
 \begin{table}[!ht]
 \renewcommand{\thetable}{S\arabic{table}}
\centering
\begin{tabular}{ccc}
 Polystyrene &MEH-PPV&MEH-PPV / Polystyrene gels\\
 \hline
    Concentration(mg/ml) &Molarity (M)&(mole percent$\%$)$\times10^{-3}$\\
  \hline
 15\,mg/ml &5.0$\times10^{-5}$\,M& 3.5\\
 30\,mg/ml &5.0$\times10^{-5}$\,M & 1.7\\
 50\,mg/ml&5.0$\times10^{-5}$\,M&1.04\\
 50\,mg/ml&5.0$\times10^{-4}$\,M&1.04\\
 100\,mg/ml&5.0$\times10^{-5}$\,M &0.52\\
  \hline
\end{tabular}
\caption{Characteristics of different  gels}
\end{table}

\textbf{ Calculation of the exciton coherence length}
In a linear aggregate the coherence length is given by,
\begin{equation}
L_{c}=\left ( N_{coh} -1\right )d,
\end{equation}
where $d$ is the distance between nearestneighbour chromophores. The coherence number at $T=0$\,K in aggregates with no disorder  corresponds to $L_{c}=\left ( N_{coh} -1\right )d$, therefore $N_{coh}=\frac{L_{c}}{d}+1\sim 6$, where $\frac{L_{c}}{d}$ is taken from Fig. 8.
\newpage
\textbf
{Resonance Raman Spectroscopy}
 \begin{figure}[H]
 \centering
 \renewcommand{\thefigure}{S\arabic{figure}}
 \includegraphics{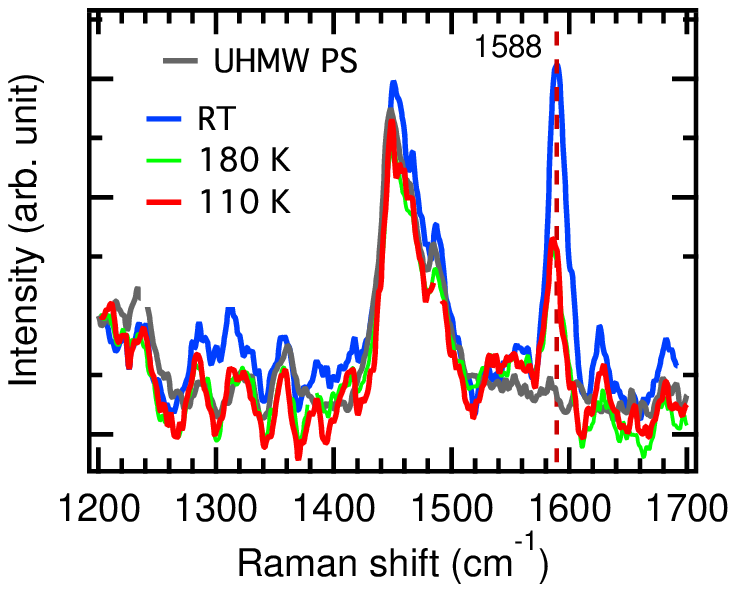}
 \caption{ Temperature dependent resonance Raman spectra of MEH-PPV in UHMW PS/MeTHF pseudogels. The spectra are normalised with respect to the 1448\,cm$^{-1}$ band of MeTHF. The dashed lines indicate spectral changes.}
 \label{fig:gel-raman}
 \end{figure}


